\documentclass[conference]{IEEEtran}
\IEEEoverridecommandlockouts
\usepackage{cite}
\usepackage{amsmath,amssymb,amsfonts}
\usepackage{algorithmic}
\usepackage{graphicx}
\usepackage{textcomp}
\usepackage{xcolor}
\usepackage[symbol]{footmisc}

\def\BibTeX{{\rm B\kern-.05em{\sc i\kern-.025em b}\kern-.08em
    T\kern-.1667em\lower.7ex\hbox{E}\kern-.125emX}}
\begin{document}

\title{A Critical Appraisal of Data Augmentation Methods for Imaging-Based Medical Diagnosis Applications}

\author{
    \IEEEauthorblockN{
    Tara M. Pattilachan\IEEEauthorrefmark{1}\textsuperscript{\textparagraph}, 
    Ugur Demir\IEEEauthorrefmark{2}\textsuperscript{\textparagraph}, 
    Elif Keles\IEEEauthorrefmark{2},
    Debesh Jha\IEEEauthorrefmark{2},
    Derk Klatte\IEEEauthorrefmark{3},
    Megan Engels\IEEEauthorrefmark{3},\\
    Sanne Hoogenboom\IEEEauthorrefmark{3},
    Candice Bolan\IEEEauthorrefmark{3},
    Michael Wallace\IEEEauthorrefmark{4}, 
    Ulas Bagci\IEEEauthorrefmark{2}}
    \IEEEauthorblockA{\IEEEauthorrefmark{1}University of Central Florida, FL, USA,\\
\IEEEauthorrefmark{2}Machine and Hybrid Intelligence Lab, Northwestern University, IL, USA,\\
\IEEEauthorrefmark{3} Mayo Clinic, Jacksonville, FL, USA,\\
\IEEEauthorrefmark{4} Mayo Clinic,  Sheikh Shakhbout Medical City, UAE}
}

\maketitle
\begingroup\renewcommand\thefootnote{\textparagraph}
\endgroup

\begin{abstract}
Current data augmentation techniques and transformations are well suited for improving the size and quality of natural image datasets but are not yet optimized for medical imaging. We hypothesize that sub-optimal data augmentations can easily distort or occlude medical images, leading to false positives or negatives during patient diagnosis, prediction, or therapy/surgery evaluation. In our experimental results, we found that utilizing commonly used intensity-based data augmentation distorts the MRI scans and leads to texture information loss, thus negatively affecting the overall performance of classification. Additionally, we observed that commonly used data augmentation methods cannot be used with a plug-and-play approach in medical imaging, and requires manual tuning and adjustment.


\end{abstract}


\section{Introduction}

In this study, we investigate how commonly used data augmentation methods affect medical imaging based diagnosis tasks, the following methods are used: RandAugment~\cite{randaug}, AutoAugment~\cite{aa}, Fast AutoAugment~\cite{faa}, Trivial Augment~\cite{trivialaug} and AugMix~\cite{augmix}.

\begin{figure}[!h]
\centerline{\includegraphics[width=0.8\linewidth]{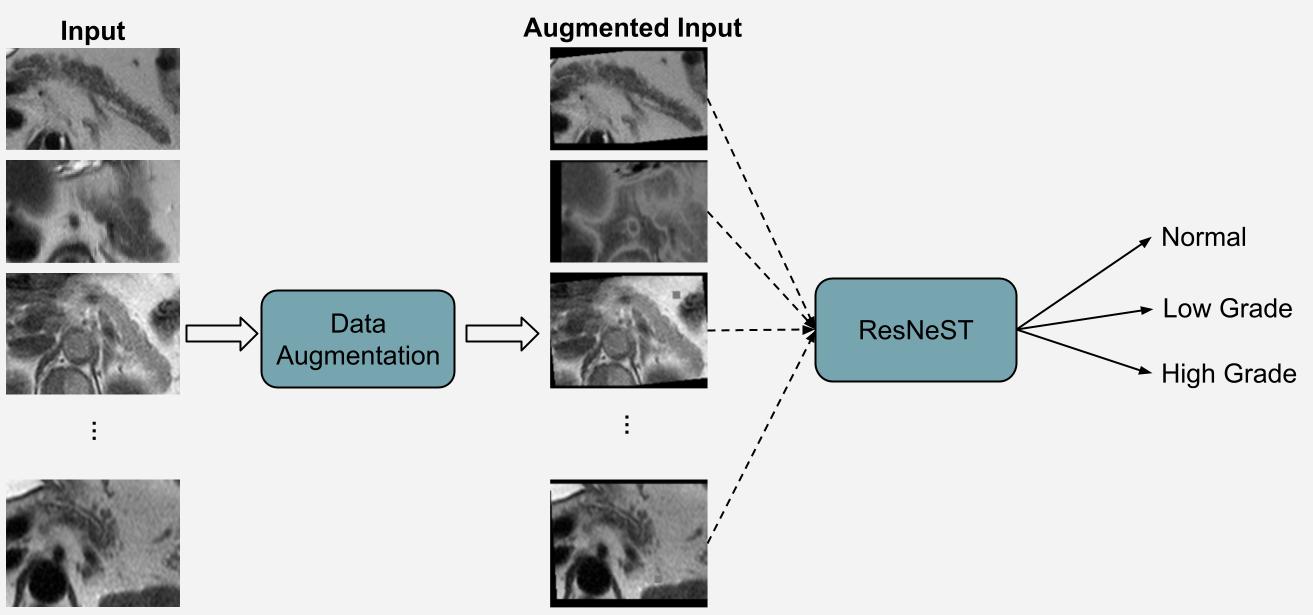}}
\vspace{-3mm}
\caption{Overall architecture for ResNeST~\cite{zhang2020resnest} training with data augmentation.}
\label{fig:archicture}
\vspace{-6mm}
\end{figure}

 \section{Method}
 Figure~\ref{fig:archicture} shows the block diagram of the proposed method. 
In this study, we tested commonly used data augmentation methods RandAugment \cite{randaug}, AutoAugment~\cite{aa}, Fast AutoAugment~\cite{faa}, Trivial Augment~\cite{trivialaug} and AugMix~\cite{augmix}, and their impact on the MRI based IPMN classification problem. The classification problem was modelled via a deep learning architecture, called ResNeST~\cite{zhang2020resnest}, where we classified the images into normal, low grade and high grade categories.

\begin{table}
\caption{MRI based IPMN classification results with respect to data augmentation methods.}.
\vspace{-7mm}
\begin{center}
\begin{tabular}{l|c}
\hline
\textbf{Method} & \textbf{Accuracy} \\
\hline
Baseline & $61.70\% \pm 9.42$ \\
\hline
RandAug & $55.30\% \pm 9.29$ \\
\hline
RandAug - geometric & $61.67\% \pm 8.37$ \\
\hline
Auto Augment &  $51.03\% \pm 10.31$ \\
\hline
Fast Auto Augment &  $55.34\% \pm 12.78$ \\
\hline
Trivial Augment &  $51.75\% \pm 9.30$ \\
\hline
AugMix & $54.56\% \pm 8.19$ \\
\hline
\end{tabular}
\label{table:results}
\end{center}
\vspace{-7mm}
\end{table}

\subsection{Results}

Table~\ref{table:results} exhibits the performance of each augmentation method and the baseline. It can be observed that the baseline achieves an accuracy of $61.70\% \pm 9.42$. However, other augmentation methods such as RandAug, Auto Augment, Fast Auto Augment, Trivial Augment and AugMix are causing significant performance drop (refer Table~\ref{table:results}). Table shows that when the intensity based augmentation is removed performance stays close to the baseline in the ``RandAug - geometric" experiments where it achieves $61.67\% \pm 8.37$.

\section{Conclusion}
In this study, we raised critical appraisals for the role of data augmentation for medical imaging tasks. We analyzed five commonly used data augmentation approaches and their effect on the performance of the MRI based IPMN classification problem. Our study in the controlled experiments showed that the commonly used data augmentation methods are designed  specifically for natural images and they can have adverse effects in medical diagnosis tasks if used without modification.



 \subsubsection*{Acknowledgement}
 This project is supported by the NIH funding: R01-CA246704 and R01-CA240639.  
\vspace{-2mm}
\bibliographystyle{IEEEtran}
\bibliography{mybib}

\end{document}